\documentstyle[12pt,a4]{article}

\newcommand{\news}{\setcounter{equation}{0}}

\newcommand{\grad}{\mbox{$\bigtriangledown$}}
\def\eqn{\begin{equation}}
\def\eeqn{\end{equation}}
\def\arr{\begin{array}}
\def\earr{\end{array}}
\def\a{\alpha}

\def\e{\epsilon}
\def\o{\Omega ^{2}}
\def\G{\Gamma}

\begin{document}

\title{Composite M-branes}

\author{Miguel S. Costa\thanks{email address:
M.S.Costa@damtp.cam.ac.uk}\\\\
D.A.M.T.P.\\
University of Cambridge\\
Silver Street\\
Cambridge CB3 9EW\\
England}

\date{September 1996}

\maketitle

\begin{abstract}
We present new supersymmetric solutions of $D=11$ supergravity
obtained by intersecting the brane configuration interpreted as a
2-brane lying within a 5-brane. Some of these solutions can be boosted
along a common string and/or  superposed with a Ka\l u\.{z}a-Klein
monopole. We also present a new embedding of the extreme four
dimensional dyonic black hole with finite horizon area. These
solutions are a consequence of a rather simple set of rules that allow
us to construct the composite M-branes.
\end{abstract}

\newpage

\section{Introduction}\
\news

Recently there has been a renewed interest in studying eleven
dimensional supergravity brane solutions. This is due to a conjectured
quantum theory in eleven dimensions, M-theory, whose effective field
theory limit is $11D$ supergravity \cite{t,w}. The latter has electric
membrane \cite{ds} and magnetic 5-brane \cite{g} solutions with
charges arising from the 4-form field strength of this supergravity
theory. These extreme solutions preserve half of the maximal
supersymmetries and are believed to play a crucial role in a precise
formulation of M-theory. In particular, the supermembrane admits a
covariant Green-Schwartz action \cite{bst}. Double dimensional
reduction of the supermembrane action from eleven to ten dimensions
gives rise to the type IIA superstring action \cite{d}. This is a
strong evidence for some relation between an underlying quantum theory
in eleven dimensions and type IIA superstring theory. The conjectured
M-theory is believed to yield upon compactification the different
superstring theories:
compactification on $S^{1}$ gives rise to type IIA superstring theory
\cite{w}; the $E_{8}\times E_{8}$ heterotic string is believed to
arrive in the reduction on $\frac{S^{1}}{Z_{2}}$ \cite{hw}; the type
II and the heterotic-type I dualities are related to the
compactification of the M-theory on $T^{2}$ and on a
$Z_{2}$ orbifold of $T^{2}$, respectively $[7-9]$.

It is known long ago that the basic electric 2-brane and magnetic
5-brane solutions of $11D$ supergravity (now called M-branes) are not
the only supersymmetric brane solutions of this theory
\cite{g}. However, Townsend and Papadopoulos \cite{pt} have
reinterpreted these other solutions as orthogonal intersections of
M-branes and found other supersymmetric orthogonal intersecting
solutions. Tseytlin \cite{ts} (see also \cite{kt,gkt}) has extended
this work and formulate a general rule, the `harmonic function rule',
to construct orthogonal intersecting branes with the basic 2 and
5 branes as its constituents.  Some configurations can be boosted
along a common string to all branes and/or superposed with a Ka\l
u\.{z}a-Klein (KK) monopole \cite{gp,so} yielding, upon dimensional
reduction to $D=10$, brane solutions with charges arising from the R-R
2-form field strength of type IIA superstring theory. There is
however, another configuration of M-branes interpreted as a 2-brane
lying within a 5-brane ($(2\subset 5)$-brane) corresponding to the
uplift to eleven dimensions of the dyonic membrane of $N=2$, $D=8$
supergravity \cite{ilpt}. It seems therefore natural that there should
be intersecting configurations with the previous solution among its
constituents. The aim of this paper is to present new intersecting
solutions of $(2\subset 5)$-branes and to formulate general rules to
construct the composite M-branes. Building all these M-brane
configurations and deducing their composite rules is important not
only to understand what M-theory may or may not be, but also to
find new the supersymmetric brane solutions of string theories as
these can be obtained from reductions of the composite M-branes
and further duality transformations $[11,13,17-19]$. An
interesting application is to perform the
statistical counting of the entropy of extreme (or near extreme) black
holes from an eleven dimensional perspective $[12,20-22]$.

This paper is organised as follows. In section two we set our notation
and state the rules to construct the composite M-branes. In section
three we present the new solutions: $(2\subset 5)\perp (2\subset
5)\perp (2\subset 5)$, $(2\subset 5)\perp 5+boost$, $(2\subset
5)\perp 2+KK\ monopole$, $(2\subset 5)+boost+KK\ monopole$ and $2\perp
2\perp 2+KK\ monopole$ branes. The latter is an embedding of the
extreme four dimensional dyonic black hole with finite horizon
area $[23-25]$. In section four we present a table with the composite
M-brane scan and give some concluding remarks.
\

\section{Composite M-brane rules}\
\news

The action for the bosonic sector of eleven dimensional supergravity
is
\eqn
I_{11}=\frac{1}{2}\int
d^{11}x\sqrt{-g}\left[R-\frac{1}{2.4!}{{\cal F}}^{2}\right]+
\frac{1}{12}\int {\cal F}\wedge {\cal F}\wedge {\cal A},
\eeqn
where ${\cal F}=d{\cal A}$ and ${\cal A}$ is a 3-form
field. The equations of motion that follow from this action are
\eqn
\arr{l}
R_{MN}=\frac{1}{2.3!}\left({\cal F}_{MPQR}{\cal
  F}_{N}^{\ \ PQR}-\frac{1}{12}g_{MN}{\cal F}^{2}\right),
\\\\ \partial _{M}\left( \sqrt{-g}{\cal F}^{MNPQ} \right)
+\frac{1}{2.4!4!}\e ^{NPQR_{1}...R_{8}}{\cal
    F}_{R_{1}...R_{4}}{\cal F}_{R_{5}...R_{8}}=0,
\earr
\eeqn
where $\e$ is the alternating symbol defined by $\e
^{tx_{1}...x_{10-p}y_{1}...y_{p}}=1$ with $y_{i}$ the spatial
coordinates of the anisotropic $p$-brane and $x_{i}$ the remaining
spatial coordinates. A given bosonic background is said to be
supersymmetric if there is a Killing spinor field solving the equation
\eqn
\delta \Psi _{M}=\left[ D_{M}+\frac{1}{288}\left(
\G_{M}^{\ \ NPQR}-8\delta_{M}^{N}\G^{PQR}\right){\cal F}_{NPQR}\right]\e
=0,
\eeqn
where we use the gamma matrices algebra
$\{\G_{A},\G_{B}\}=2\eta_{AB}$. As usual $\G_{M}=e_{M}^{\ \ A}\G_{A}$,
where $M,N,...$ denote world indices and $A,B,...$ tangent space
indices.\footnote{We will order the world indices in $e_{M}^{\ \
    A}$ according to $ty_{1}...y_{p}x_{1}...x_{10-p}$.}

Now we define the relative transverse space to a given
$q_{i}$-brane. A $q_{1}$-brane intersecting a $q_{2}$-brane
over a $r$-brane can be seen as an anisotropic
($p=q_{1}+q_{2}-r$)-brane, and further on for configurations with more
than two intersecting branes. The tangent vectors to the basic
$q_{i}$-branes' worldvolume that are not tangent to the $r$-brane's
worldvolume span the relative transverse space. The overall transverse
space is spanned by the vectors orthogonal to the $r$-brane's
worldvolume and to the relative transverse space \cite{pt}. The
relative transverse space to a given $q_{i}$-brane, denoted by ${\cal
  M}_{(q,i)}^{p-q}$, is the space spanned by the vectors that are
orthogonal to the $q_{i}$-brane's worldvolume and that are tangent to
the anisotropic p-brane's worldvolume.\footnote{If we have a
  configuration with the harmonic functions independent of $n$
  directions of the overall transverse space then these $n$ directions
  belong to the anisotropic $p$-brane worldvolume, e.g. a 2-brane will
  be seen as an anisotropic ($p=2+n$)-brane. These $n$ directions of
  the overall transverse space also belong to the branes' relative
  transverse spaces. The corresponding coordinates will be denoted by
  $y_{i}$.} 

\subsection{Orthogonal intersections of 2 and 5 branes}\

The basic $\frac{1}{2}$ supersymmetric 2-brane solution of $11D$
supergravity is described by ($2\le p\le 7$)
\eqn
\arr{l}
ds^{2}=H^{\frac{1}{3}}\left[ H^{-1}\left(
-dt^{2}+dy_{1}^{2}+dy_{2}^{2}\right) +dy_{3}^{2}+...+dy_{p}^{2}+
dx^{i}dx_{i}\right],
\\\\ **{\cal F}_{(2)}=Q\left(\e _{9-p}\wedge\eta\right),
\earr
\eeqn
where $H=1+\frac{\a}{r^{8-p}}$ is a harmonic function on the overall
transverse space and $r^{2}=x^{i}x_{i}$ with $i=1,...,10-p$. $\eta
=dy_{3}\wedge ...\wedge dy_{p}$ is the volume form on
the 2-brane relative transverse space ${\cal M}_{(2)}^{p-2}$ and $\e
_{9-p}$ the unit ($9-p$)-sphere volume form. The electric charge is
defined by $Q=\frac{1}{V_{(2)}^{p-2}}\int _{\Sigma}*{\cal F}_{(2)}$,
where $V_{(2)}^{p-2}$ is the volume of ${\cal M}_{(2)}^{p-2}$ and
$\Sigma =S^{9-p}\times {\cal M}_{(2)}^{p-2}$ is an asymptotic
spacelike hypersurface.\footnote{The Hodge dual of a $n$-form ${\cal
    F}$ in a $D$ dimensional manifold with Lorentzian signature is
  defined by $\left(*{\cal F}\right)_{B_{1}...B_{D-n}}=
  \frac{1}{n!}\sqrt{-g}\e_{A_{1}...A_{n}B_{1}...B_{D-n}} {\cal
    F}^{A_{1}...A_{n}}$ and satisfies $*^{2}=(-1)^{1+n(D-n)}$. We
  define $\e_{tx_{1}...x_{D-1-p}y_{1}...y_{p}}=-1$ such that
  $\e^{A_{1}...A_{D}}\e_{A_{1}...A_{D}}=-D!$. We will always define
  the contribution to ${\cal F}$ of an electric charged basic
  ($q=n-2$)-brane such that ${\cal F}_{(q)_{try_{i}...y_{i+q}}}=
  \frac{Q}{A_{D-2-p}}\frac{H^{-2}}{r^{D-2-p}}$ for some $i$.} 
The charge $Q$ is related to the positive constant $\a$ by
\eqn
\frac{Q}{(8-p)A_{9-p}}=\pm\a ,
\eeqn
where $A_{9-p}$ is the volume of the unit ($9-p$)-sphere. 

This solution has a Killing spinor field given by
\eqn
\e =H^{-\frac{1}{6}}\e _{0},
\eeqn
where $\G_{012}\e_{0}=\mp\e_{0}$. The $\mp$ sign choice corresponds to
positive or negative charge, respectively. This 
solution can be generalised to the corresponding multi-centered
cases. In order to keep the notation less cumbersome we will consider
just brane solutions centered at the origin.
 
The other $\frac{1}{2}$ supersymmetric basic solution is the magnetic
5-brane described by ($5\le p \le 7$)
\eqn
\arr{l}
ds^{2}=H^{\frac{2}{3}}\left[ H^{-1}\left(
-dt^{2}+dy_{1}^{2}+...+dy_{5}^{2}\right) +dy_{6}^{2}+...+dy_{p}^{2}+
dx^{i}dx_{i}\right],
\\\\ {\cal F}_{(5)}=P\left(\mu\wedge\e _{9-p}\right),
\earr
\eeqn
where $H=1+\frac{\a}{r^{8-p}}$ is a harmonic function on the overall
transverse space, $r^{2}=x^{i}x_{i}$ with $i=1,...,10-p$ and
$\mu =dy_{6}\wedge ...\wedge dy_{p}$ is the volume form on ${\cal
  M}_{(5)}^{p-5}$. The magnetic charge is defined by
$P=\frac{1}{V_{(5)}^{p-5}}\int _{\Sigma}{\cal F}_{(5)}$, where
$V_{(5)}^{p-5}$ is the volume of ${\cal M}_{(5)}^{p-5}$ and $\Sigma
={\cal M}_{(5)}^{p-5}\times S^{9-p}$. The charge $P$ is also given by
the formula (2.5). 

The Killing spinor field is
\eqn
\e =H^{-\frac{1}{12}}\e _{0},
\eeqn
where $\G_{678910}\e_{0}=\mp\e_{0}$. The $\mp$ sign choice
corresponds to positive or negative charge, respectively. \footnote{We
  order the gamma matrices indices such that they grow from the left
  to the right. This way there is no confusion between
  $\G_{01}=\G_{[0}\G_{1]}$ and $\G_{10}$. The latter is defined by
  $\G_{10}=\G^{10}=\G^{0}\G^{1}...\G^{9}$.}

The previous basic branes can be used to construct supersymmetric
configurations of N branes intersecting each other orthogonally
\cite{pt,ts}. These configurations may be seen as anisotropic
$p$-branes and can be built according to the following rules:

\begin{description}
\item[(i)]
To each basic $q_{i}$-brane we assign an harmonic function $H_{i}$ on the
overall transverse space. If the coordinate $y$ belongs to several
constituents $q_{i}$-branes ($q_{1},...,q_{n}$) then its contribution to
the metric written in the conformal frame where the overall transverse
space is `free' is $H_{1}^{-1}...H_{n}^{-1}dy^{2}$
\cite{ts}.\footnote{There is an extra minus sign if $y$ is the time
  coordinate $t$.} The contribution to the conformal factor of the
i-th $q_{i}$-brane is $H_{i}^{\frac{q_{i}+1}{9}}$. The 4-form field
strength is given by
\eqn
{\cal F}=\sum _{i=1}^{N} {\cal F}_{(q,i)},
\eeqn
where
\eqn
*{\cal F}_{(2,i)}=Q_{i}(\e_{9-p}\wedge\eta_{i}),\ \ 
{\cal F}_{(5,i)}=P_{i}(\mu_{i}\wedge \e_{9-p}),
\eeqn
whether the i-th brane is a 2 or 5-brane, respectively. In the former
case, $\eta_{i}$ is the volume form on the i-th 2-brane relative
transverse space ${\cal M}_{(2,i)}^{p-2}$ and the electric charge is defined
by $Q_{i}=\frac{1}{V_{(2,i)}^{p-2}}\int _{\Sigma}*{\cal F}_{(2,i)}$ with
$\Sigma =S^{9-p}\times {\cal M}_{(2,i)}^{p-2}$ and $V_{(2,i)}^{p-2}$
the volume of ${\cal  M}_{(2,i)}^{p-2}$. In the latter case, $\mu_{i}$
is the volume form on ${\cal M}_{(5,i)}^{p-5}$ and the magnetic charge is
defined by $P_{i}=\frac{1}{V_{(5,i)}^{p-5}}\int _{\Sigma}{\cal F}_{(5,i)}$
with $\Sigma ={\cal M}_{(5,i)}^{p-5}\times S^{9-p}$ and $V_{(5,i)}^{p-5}$
the volume of ${\cal  M}_{(5,i)}^{p-5}$. Both the electric and
magnetic charges are given by
\eqn
\frac{Q_{i}}{(8-p)A_{9-p}},\ \frac{P_{i}}{(8-p)A_{9-p}}=\pm\a_{i}.
\eeqn
\

\item[(ii)]
$q$-branes of the same type can intersect orthogonally over
($q-2$)-branes \cite{pt}. A 2-brane can intersect orthogonally a
5-brane over a string \cite{st,t2}.

\end{description}

\subsection{Intersections with the ($2\subset 5$)-brane}\

Consider now the M-brane configuration interpreted as a 2-brane lying
within a 5-brane \cite{ilpt}, i.e. the ($2\subset 5$)-brane ($5\le p
\le 7$)
\eqn
\arr{l}
\arr{ll}
ds^{2}= & \left(H\tilde{H}\right)^{\frac{1}{3}}\left[ H^{-1}\left(
-dt^{2}+dy_{1}^{2}+dy_{2}^{2}\right) \phantom{\tilde{H}^{-1}}\right.
\\\\ & \left. + \tilde{H}^{-1}\left( dy_{3}^{2}+dy_{4}^{2}+
dy_{5}^{2}\right)+dy_{6}^{2}+...+dy_{p}^{2}+dx^{i}dx_{i}\right],
\earr
\\\\ {\cal F}_{(2\subset 5)}={\cal F}_{(2)}+{\cal F}_{(5)}
-(8-p)\frac{\a \sin{2\zeta}}{2}\frac{\tilde{H}^{-2}}{r^{9-p}}
\left(dr\wedge\xi\right),
\earr
\eeqn
where $*{\cal F}_{(2)}=Q(\e_{9-p}\wedge\eta)$, ${\cal
  F}_{(5)}=P(\mu\wedge\e_{9-p})$, 
$H=1+\frac{\a}{r^{8-p}}$ and $\tilde{H}=1+\frac{\tilde{\a}}{r^{8-p}}$
are harmonic functions on the overall transverse space, 
$r^{2}=x^{i}x_{i}$ with $i=1,...,10-p$ and $\tilde{\a}=\a
\cos^{2}{\zeta}$. If $\cos{\zeta}=0$ we obtain the 2-brane solution
(2.4) and if $\sin{\zeta}=0$ the 5-brane solution (2.7). $\mu
=dy_{6}\wedge ...\wedge dy_{p}$ and $\eta
=dy_{3}\wedge dy_{4} \wedge dy_{5}\wedge\mu$ are the volume forms on
the relative transverse spaces of the constituent M-branes (${\cal
  M}_{(5)}^{p-5}$ and ${\cal M}_{(2)}^{p-2}$). $\xi
= dy_{3}\wedge dy_{4}\wedge dy_{5}$ is the volume form on the space
${\cal M}_{(5/2)}^{3}$ spanned by the vectors that are tangent to the
5-brane's worldvolume but are orthogonal to the 2-brane's
worldvolume. The dual operation in the expression for ${\cal F}_{(2)}$
is defined for the metric with $\tilde{\a}=0$. The electric and
magnetic charges are defined as for the previous basic solutions and
they are given by \footnote{We remark that from the 4-form field
  strength equation in (2.2) the conserved electric charge is in fact
  given by $Q=\frac{1}{V_{(2)}^{p-2}}\int_{\Sigma}\left( *{\cal
      F}+\frac{1}{2}{\cal A}\wedge{\cal F}\right)$. Even thought the
  Chern-Simons term does not vanish for this solution, we can always
  choose a gauge such that it vanishes at spatial infinity and
  therefore the electric charge may still be defined as before.}
\eqn
\frac{Q}{(8-p)A_{9-p}}=\a\sin{\zeta}\ ,\ \ 
\frac{P}{(8-p)A_{9-p}}=\a\cos{\zeta}.
\eeqn
This background preserves half of the supersymmetries, admitting the
following Killing spinor field
\eqn
\e =\left( H\tilde{H} \right)^{-\frac{1}{6}}\left[
\left(\tilde{H}^{\frac{1}{2}}\pm H^{\frac{1}{2}}\cos{\zeta}\right)
  ^{\frac{1}{2}}+\left(\tilde{H}^{\frac{1}{2}}\mp H^{\frac{1}{2}}
  \cos{\zeta}\right)^{\frac{1}{2}} \gamma \right] \e_{0},
\eeqn
where
\eqn
\G_{012} \gamma \e_{0}=\mp \e_{0},
\eeqn
with $\gamma=\G_{012}\G_{678910}$. The upper signs choice is for
$\sin{\zeta}\ge 0$ and the lower signs choice for $\sin{\zeta}\le
0$. If $\cos{\zeta}=0$ or $\sin{\zeta}=0$ we obtain Killing spinor
fields that can be written as in (2.6) or (2.8), respectively.

Configurations of intersecting M-branes including the ($2\subset
5$)-branes as possible constituents can be constructed according to
the rule:

\begin{description}
\item[(iii)]
If there are ($2\subset 5$)-branes among the $N$ intersecting
constituents then in the cases where these branes reduce to the basic 2
or 5-branes the resulting configuration should be compactible with the
previous rules (i) and (ii). The 4-form field strength is given by
\eqn
{\cal F}=\sum _{i=1}^{N} {\cal F}_{(q,i)},
\eeqn
where ${\cal F}_{(q,i)}$ is given either by one of the expressions in
(2.10) or by\footnote{The sign of the third term in (2.17) depends on
  how we distribute the constituent M-branes. We will always make the
  minus sign choices.}
\eqn
{\cal F}_{(2\subset 5,i)}={\cal F}_{(2,i)}+{\cal F}_{(5,i)}-(8-p)
\frac{\a_{i}\sin{2\zeta_{i}}}{2}\frac{\tilde{H}_{i}^{-2}}{r^{9-p}}
\left(dr\wedge \xi_{i}\right) .
\eeqn
The dual operations are defined for $\tilde{\a}_{i}=0$ and $\xi_{i}$
is the volume form on the space ${\cal M}_{(5/2,i)}^{3}$. The
electromagnetic charges are defined as before. They are given by
\eqn
\frac{Q_{i}}{(8-p)A_{9-p}}=\a_{i}\sin{\zeta_{i}}\ ,\ \ 
\frac{P_{i}}{(8-p)A_{9-p}}=\a_{i}\cos{\zeta_{i}}\ ,
\eeqn
where $\cos{\zeta_{i}}=0$ or $\sin{\zeta_{i}}=0$ if the i-th brane is
a basic 2 or 5-brane, respectively.

\end{description}

\subsection{Ka\l u\.{z}a-Klein charges}\

In some configurations we can add KK charges yielding, upon
dimensional reduction to $D=10$, branes with charges arising from the
R-R 2-form field strength of type IIA superstring theory.
The Ka\l u\.{z}a-Klein reduction of $11D$ supergravity yields $N=2A$,
$D=10$ supergravity. The corresponding bosonic sector action written
in the string frame is given by
\eqn
\arr{ll}
I_{IIA}= & \frac{1}{2}\int d^{10}x\sqrt{-g}
        \left[ e^{-2\phi}\left(R+4(\grad \phi)^{2}
        -\frac{1}{2.3!}{\cal H}^{2}\right)
        -\frac{1}{2.2!}{\cal F}_{2}^{2} 
        -\frac{1}{2.4!}{\cal F}_{4}'^{2}\right]
\\\\ & +\frac{1}{4}\int {\cal F}_{4}\wedge {\cal F}_{4}\wedge {\cal B},
\earr
\eeqn
where ${\cal F}_{4}'=d{\cal A}_{3}+{\cal A}\wedge {\cal H}$, ${\cal
  F}_{4}=d{\cal A}_{3}$, ${\cal H}=d{\cal B}$, ${\cal F}_{2}=d{\cal A}$
and ${\cal A}$, ${\cal B}$ and ${\cal A}_{3}$ are 1, 2 and 3-form fields,
respectively. The reduction of the bosonic fields is performed by
writing
\eqn
\arr{l}
ds^{2}=g_{MN}dx^{M}dx^{N}=e^{-\frac{2}{3}\phi
}g_{mn}dx^{m}dx^{n}+e^{\frac{4}{3}\phi
}\left( dx^{10}-{\cal A}_{m}dx^{m}\right)^{2},\\\\
{\cal A}_{MNP}={\cal A}_{mnp},\ {\cal A}_{MN10}={\cal B}_{mn},\ \ with\
\ M,N,P=0,...,9.
\earr
\eeqn
Unless stated capital letters range from 0 to 10 and in the lower case
from 0 to 9. The rule to generate branes with KK charges is:

\begin{description}
\item[(iv)]
If there is a common string to all constituent M-branes, say along
$y$, then an electric KK charge can be added by applying a boost along
the $y$ direction \cite{ts}

\eqn
-dt^{2}+dy^{2}\rightarrow -dt^{2}+dy^{2}+
\frac{\a}{r^{8-p}}(dy\mp dt)^{2},
\eeqn
where the $\mp$ sign choice will correspond to positive or negative KK
charge, respectively. Compactifying along the $y$ direction we have a
collection of branes intersecting a 0-brane over a point. The 2-form
field strength is then given by $(p\rightarrow p-1)$
\eqn
*{\cal F}_{2}=Q\left(\e_{8-p}\wedge\eta\right),
\eeqn
where $\eta$ is the volume form of the relative transverse space of
the 0-brane ${\cal M}_{(0)}^{p}$. The KK electric charge is defined by
$Q=\frac{1}{V_{(0)}^{p}}\int _{\Sigma}*{\cal F}_{2}$ with $V_{(0)}^{p}$
the volume of  ${\cal M}_{(0)}^{p}$ and $\Sigma=S^{8-p}\times {\cal
  M}_{(0)}^{p}$. This charge is given by
\eqn
\frac{Q}{(7-p)A_{8-p}}=\pm\a.
\eeqn

If the overall transverse space has dimension bigger than
three, a magnetic monopole can be added by making all the harmonic
functions to depend only on three of this space coordinates and
performing the substitution (we start with an anisotropic $p$-brane)
\eqn
\arr{ll}
dx^{i}dx_{i}\rightarrow & dy_{p+1}^{2}+...+dy_{6}^{2} 
\\\\ & +H^{-1}\left( dy_{7}\pm \a \cos{\theta}d\phi \right)^{2}+
H\left(dr^{2}+r^{2}d\o _{2}\right),
\earr
\eeqn
where we have introduced spherical coordinates, $H=1+\frac{\a}{r}$ and
the $\pm$ sign choice will correspond to positive or negative KK
charge, respectively. Compactifying along the
$y_{7}$ direction we have a collection of branes superposed over a
6-brane. The 2-form field strength is then given by
\eqn
{\cal F}_{2}=P\e_{2},
\eeqn
where the KK magnetic charge is defined by
$P=\int _{S^{2}}{\cal F}_{2}$ and is given by
\eqn
\frac{P}{A_{2}}=\pm\a.
\eeqn

There are cases where we can add both electric and magnetic KK
charges and a further compactification is required.

\end{description}

\subsection{Supersymmetry}\

The previous rules yield supersymmetric backgrounds. The amount of
unbroken supersymmetries is determined by the rule:

\begin{description}
\item[(v)]
A configuration with n constituent M-branes and KK charges preserves
$2^{-n}$ of the maximal supersymmetry. \footnote{In the cases of
  multi-centered solutions, a given family of parallel branes
  contributes only once to n.} There are exceptions to this
rule. Namely, the embeddings of the four dimensional dyonic black hole
with finite horizon area have $n=4$ but preserve $\frac{1}{8}$ of the
maximal supersymmetry or break all supersymmetries. This happens
because the conditions on the Killing spinor field due to each of the
constituent branes are not independent or are incompactible,
respectively.

\end{description}

The fact that we considered supersymmetric backgrounds means that they
will satisfy a Bogolmol'nyi bound for the ADM mass
$[16,28-31]$. The ADM mass per unit of $p$-volume for an
anisotropic $p$-brane in D dimensions described by the Einstein's
frame metric
\eqn
ds^{2}=-A(r)dt^{2}+\sum_{k=1}^{p}B_{k}(r)dy_{k}^{2}+C(r)\left(
dx_{1}^{2}+...+dx_{D-p-1}^{2}\right),
\eeqn
is given by a straightforward generalisation of the result presented
in \cite{lu}
\eqn
\frac{M}{A_{D-2-p}}=-\frac{1}{2}\left[r^{D-2-p}\partial_{r}\left(
\sum_{k=1}^{p}B_{k}(r)+(D-2-p)C(r)\right) \right]_{r\rightarrow +\infty}.
\eeqn
We wrote this result in D dimensions in order to accommodate the cases
with electric or magnetic (or both) KK charges. Labelling all the
charges by $Q_{i}$ (electric, magnetic and KK in origin) and defining
in the ($2\subset 5$)-brane case an electromagnetic charge $Q_{i}$ by
$Q_{i}^{2}=P_{2\subset 5,i}^{2}+Q_{2\subset 5,i}^{2}$\ , we have that
the composite M-brane rules imply
\eqn
2M=\sum_{i=1}^{n}|Q_{i}|.
\eeqn

\section{New composite M-branes}\
\news

In this chapter we present the new supersymmetric solutions that
follow from applying the composite M-brane rules (iii) and (iv). By
making some of the charges to vanish it is possible to enlarge the
overall transverse space dimension of some configurations and
therefore to change the harmonic functions dependence. We will only
present the cases for which $p=7$. The other cases are given in the
table presented in the conclusion.

\subsection{$(2\subset 5)\perp (2\subset 5)\perp (2\subset 5)$ brane}\

The most general intersecting configuration of M-branes involving the
($2\subset 5$)-brane is obtained by intersecting 3 of these composite
M-branes. This solution is described by
\eqn
\arr{l}
\arr{ll}
ds^{2}= & \left({\displaystyle \prod_{i=1}^{3}}\left( H_{i}\tilde{H}_{i}
\right)^{\frac{1}{3}}\right) \left[ -\left( H_{1}H_{2}H_{3}\right)
^{-1}dt^{2}+(H_{1}\tilde{H}_{3})^{-1}dy_{1}^{2}\right.
\\\\ & +(H_{1}\tilde{H}_{2})^{-1}dy_{2}^{2}+
(H_{2}\tilde{H}_{1})^{-1}dy_{3}^{2}+
(H_{2}\tilde{H}_{3})^{-1}dy_{4}^{2} 
\\\\ & \left. +(H_{3}\tilde{H}_{2})^{-1}dy_{5}^{2}
+(H_{3}\tilde{H}_{1})^{-1}dy_{6}^{2}+(\tilde{H}_{1}\tilde{H}_{2}
\tilde{H}_{3})^{-1}dy_{7}^{2}+dx^{j}dx_{j}\right],
\earr
\\\\
{\cal F}={\displaystyle \sum_{i=1}^{3}}\left( {\cal F}_{(2,i)}+
  {\cal F}_{(5,i)}-\frac{\a_{i}\sin{2\zeta_{i}}}{2}
\frac{\tilde{H}_{i}^{-2}}{r^{2}}(dr\wedge \xi_{i})\right),
\earr
\eeqn
where $*{\cal F}_{(2,i)}=Q_{i}(\e_{2}\wedge\eta_{i})$,
${\cal F}_{(5,i)}=P_{i}(\mu_{i}\wedge \e_{2})$,
$H_{i}=1+\frac{\a_{i}}{r}$,
$\tilde{H}_{i}=1+\frac{\tilde{\a}_{i}}{r}$,
$\tilde{\a}_{i}=\a_{i}\cos^{2}{\zeta_{i}}$ and $j=1,2,3$.
The dual operations are defined for $\tilde{\a}_{i}=0$. $\eta_{i}$,
$\mu_{i}$ and $\xi_{i}$ are the volume forms on ${\cal
  M}^{p-2}_{(2,i)}$, ${\cal M}^{p-5}_{(5,i)}$ and ${\cal
  M}^{3}_{(5/2,i)}$, respectively. The electric and magnetic charges
are defined as before and are given by (2.18) with $p=7$. The
ADM mass per unit of 7-volume obtained from (2.29) is
\eqn
2M={\displaystyle \sum_{i=1}^{3}}\sqrt{Q^{2}_{i}+P^{2}_{i}}.
\eeqn

This background admits the following killing spinor field
\eqn
\e =\left( \prod_{i=1}^{3}\left( H_{i}\tilde{H}_{i}
\right)^{-\frac{1}{6}} \left[ \left(\tilde{H}_{i}^{\frac{1}{2}}\pm 
H_{i}^{\frac{1}{2}}\cos{\zeta_{i}}\right)^{\frac{1}{2}}+
\left(\tilde{H}_{i}^{\frac{1}{2}}\mp H_{i}^{\frac{1}{2}}\cos{\zeta_{i}}
\right)^{\frac{1}{2}} \gamma_{i} \right] \right) \e_{0},
\eeqn
where
\eqn
\G_{012}\gamma_{1}\e_{0}=\mp \e_{0},\ \
\G_{034}\gamma_{2}\e_{0}=\mp \e_{0},\ \ 
\G_{056}\gamma_{3}\e_{0}=\mp \e_{0},\ \ 
\eeqn
with $\gamma_{1}=\G_{012}\G_{458910}$,
$\gamma_{2}=\G_{034}\G_{168910}$ and $\gamma_{3}=\G_{056}\G_{238910}$.
The upper and lower signs definitions are as in $(2.14-15)$. The
conditions (3.4) reduce the amount of preserved supersymmetry to 
$\frac{1}{8}$ of the maximal supersymmetry. We note that all the gamma 
matrices related with a given ($2\subset 5$)-brane commute with all 
the gamma matrices related with the other ($2\subset 5$)
branes. \footnote{On solving equation (2.3) this fact has to be
  used. For example, the gamma matrices $\gamma_{1}=-\G_{367}$,
  $\G_{012}$ and $\G_{458910}$ related with the first ($2\subset
  5$)-brane commute with all the gamma matrices related with the
  other branes.}

\subsection{Adding Ka\l u\.{z}a-Klein charges}\

We can use the composite M-brane rule (iv) to generate KK charges. The
new solutions are the $(2\subset 5)\perp 5 + boost$, $(2\subset
5)\perp 2 +KK\ monopole$, $2\perp 2\perp 2+KK\ monopole$ and
$(2\subset 5)+boost+KK\ monopole$ branes.

\subsubsection{$(2\subset 5)\perp 5 + boost$ brane}\

This solution is described by
\eqn
\arr{l}
\arr{ll}
ds^{2}= & \left( H_{1}\tilde{H}_{1}\right)^{\frac{1}{3}}
H_{2}^{\frac{2}{3}}\left[ \left( H_{1}H_{2}\right)^{-1}\left(
-dt^{2}+dy_{7}^{2}+\frac{\a_{3}}{r}(dy_{7}\mp dt)^{2}\right)\right.
\\\\ & +H_{2}^{-1}\left( dy_{1}^{2}+dy_{2}^{2}\right)+
\left(H_{2}\tilde{H}_{1}\right)^{-1}\left( dy_{3}^{2}+dy_{4}^{2}\right)
\\\\ & \left. +\tilde{H}_{1}^{-1}dy_{5}^{2}+H_{1}^{-1}dy_{6}^{2}
  +dx^{j}dx_{j}\right],
\earr
\\\\
{\cal F}={\cal F}_{(2,1)}+
\left({\displaystyle \sum_{i=1}^{2}}{\cal F}_{(5,i)}\right)
-\frac{\a_{1}\sin{2\zeta_{1}}}{2}
\frac{\tilde{H}_{1}^{-2}}{r^{2}}(dr\wedge \xi_{1}),
\earr
\eeqn
where $*{\cal F}_{(2,1)}=Q_{1}\left(\e_{2}\wedge\eta_{1}\right)$,
${\cal F}_{(5,i)}=P_{i}\left(\mu_{i}\wedge\e_{2}\right)$ 
with the dual operation defined for the metric with $\tilde{\a}_{1}=0$
and $j=1,2,3$.

This background is $\frac{1}{8}$ supersymmetric. The Killing spinor
field is 
\eqn
\e =H_{3}^{-\frac{1}{4}}H_{2}^{-\frac{1}{12}}\left( H_{1}\tilde{H}_{1}
\right)^{-\frac{1}{6}}\left[ \left(\tilde{H}_{1}^{\frac{1}{2}}\pm
H_{1}^{\frac{1}{2}}\cos{\zeta_{1}}\right)^{\frac{1}{2}}+ 
\left(\tilde{H}_{1}^{\frac{1}{2}}\mp H_{1}^{\frac{1}{2}}\cos{\zeta_{1}}
\right)^{\frac{1}{2}} \gamma \right] \e_{0},
\eeqn
where
\eqn
\G_{067}\gamma\e_{0}=\mp \e_{0},\ \
\G_{568910}\e_{0}=\mp \e_{0},\ \ 
\G_{07}\e_{0}=\mp \e_{0},\ \ 
\eeqn
with $\gamma=\G_{067}\G_{128910}$. The upper and lower sign choice in
the last condition in (3.7) corresponds to positive or negative KK
charge, respectively. 

Dimensional reduction along $y_{7}$ yields the 
$(1\subset 4)\perp 4\perp 0$ brane of type IIA superstring theory
\eqn
\arr{l}
\arr{ll}
ds_{10}^{2}= & \left( \tilde{H}_{1}H_{2}H_{3}\right)^{\frac{1}{2}}
\left[ -\left(H_{1}H_{2}H_{3}\right)^{-1}dt^{2}
 +H_{2}^{-1}\left( dy_{1}^{2}+dy_{2}^{2}\right)\right.
\\\\ & \left. 
+\left(H_{2}\tilde{H}_{1}\right)^{-1}\left( dy_{3}^{2}+dy_{4}^{2}\right)
+\tilde{H}_{1}^{-1}dy_{5}^{2}+H_{1}^{-1}dy_{6}^{2}+dx^{j}dx_{j}\right],
\earr
\\\\
e^{2\phi}=H_{1}^{-1}\tilde{H}_{1}^{\frac{1}{2}}H_{2}^{-\frac{1}{2}}
H_{3}^{\frac{3}{2}},\ \ 
*{\cal F}_{2}=Q_{3}(\e_{2}\wedge\eta_{3}),\ \ 
*{\cal H}=-Q_{1}e^{2\phi}(\e_{2}\wedge\eta_{1}),
\\\\
{\cal F}_{4}=\left({\displaystyle \sum_{i=1}^{2}}
P_{i}\left(\mu_{i}\wedge\e_{2}\right)\right)
-\frac{\a_{1}\sin{2\zeta_{1}}}{2}
\frac{\tilde{H}_{1}^{-2}}{r^{2}}(dr\wedge \xi_{1}),
\earr
\eeqn
where $\mu_{i}$,  $\eta_{1}$,  $\eta_{3}$ and $\xi_{1}$ are the
volume forms on the spaces ${\cal M}_{(4,i)}^{2}$, ${\cal
  M}_{(1,1)}^{5}$, ${\cal M}_{(0,3)}^{6}$ and ${\cal
  M}_{(4/1,1)}^{3}$, respectively. Note that according to (2.20)
dimensional reduction of ${\cal F}_{(2,1)}$ gives ${\cal
  H}_{try_{_{6}}}=\frac{Q_{1}}{A_{2}}\frac{H_{1}^{-2}}{r^{2}}$. Following
the definitions in footnote 3 the field strength ${\cal H}$ can be
written as in (3.8).

\subsubsection{$(2\subset 5)\perp 2 +KK\ monopole$ brane}\

This solution is given by
\eqn
\arr{l}
\arr{ll}
ds^{2}= & \left( H_{1}\tilde{H}_{1}H_{2}\right)^{\frac{1}{3}}
\left[ -\left( H_{1}H_{2}\right)^{-1}dt^{2}+H_{1}^{-1}\left(
  dy_{1}^{2}+dy_{2}^{2}\right)\right.  
\\\\ & +\tilde{H}_{1}^{-1}\left( dy_{3}^{2}+dy_{4}^{2}\right)+
\left(\tilde{H}_{1}H_{2}\right)^{-1}dy_{5}^{2}+H_{2}^{-1}dy_{6}^{2}
\\\\ & \left. +H_{3}^{-1}\left(dy_{7}\pm
\a_{3}\cos{\theta}d\phi\right)^{2}+H_{3}\left( dr^{2}+
r^{2}d\o_{2}\right)\right],
\earr
\\\\
{\cal F}=\left({\displaystyle \sum_{i=1}^{2}}{\cal F}_{(2,i)}\right)
+{\cal F}_{(5,1)}-
\frac{\a_{1}\sin{2\zeta_{1}}}{2}
\frac{\tilde{H}_{1}^{-2}}{r^{2}}(dr\wedge \xi_{1}),
\earr
\eeqn
where $*{\cal F}_{(2,i)}=Q_{i}\left(\e_{2}\wedge\eta_{i}\right)$ and
${\cal F}_{(5,1)}=P_{1}\left(\mu_{1}\wedge\e_{2}\right)$.

This background is $\frac{1}{8}$ supersymmetric. The Killing spinor
field is
\eqn
\e =\left( H_{1}\tilde{H}_{1}H_{2}\right)^{-\frac{1}{6}}\left[
\left(\tilde{H}_{1}^{\frac{1}{2}}\pm H_{1}^{\frac{1}{2}}
  \cos{\zeta_{1}}\right)^{\frac{1}{2}}+
  \left(\tilde{H}_{1}^{\frac{1}{2}}\mp H_{1}^{\frac{1}{2}}\cos{\zeta_{1}}
\right)^{\frac{1}{2}} \gamma \right] \e_{0},
\eeqn
where
\eqn
\G_{012}\gamma\e_{0}=\mp \e_{0},\ \
\G_{056}\e_{0}=\mp \e_{0},\ \ 
\G_{78910}\e_{0}=\pm \e_{0},\ \ 
\eeqn
with $\gamma=\G_{012}\G_{678910}$. The upper and lower sign choice in
the last condition in (3.11) corresponds to positive or negative KK
charge, respectively. 

Dimensional reduction along $y_{7}$ yields the $((2\subset 5)\perp
2)\Vert 6$ brane of type IIA superstring theory
\eqn
\arr{l}
\arr{ll}
ds_{10}^{2}= & \left( H_{1}\tilde{H}_{1}H_{2}H_{3}\right)^{\frac{1}{2}}
\left[ -\left( H_{1}H_{2}H_{3}\right)^{-1}dt^{2}\right.
\\\\ &  +\left( H_{1}H_{3}\right)^{-1}\left( dy_{1}^{2}+dy_{2}^{2}
\right) +\left( \tilde{H}_{1}H_{3}\right)^{-1}\left( dy_{3}^{2}+
dy_{4}^{2}\right)
\\\\ & \left. +\left(\tilde{H}_{1}H_{2}H_{3}\right)^{-1}dy_{5}^{2}+
\left( H_{2}H_{3}\right)^{-1}dy_{6}^{2}
+dx^{j}dx_{j}\right],
\earr
\\\\
e^{2\phi}=\left( H_{1}\tilde{H}_{1}H_{2}\right)^{\frac{1}{2}}
H_{3}^{-\frac{3}{2}},\ \ 
{\cal H}=P_{1}(\mu_{1}\wedge \e_{2})\ ,\ \ 
{\cal F}_{2}=P_{3}\e_{2}\ ,
\\\\
{\cal F}_{4}=
\left({\displaystyle \sum_{i=1}^{2}}{\cal F}_{(2,i)}\right) 
-\frac{\a_{1}\sin{2\zeta_{1}}}{2}
\frac{\tilde{H}_{1}^{-2}}{r^{2}}(dr\wedge \xi_{1}).
\earr
\eeqn

\subsubsection{$2\perp 2\perp 2 + KK\ monopole$ brane}\

There is a new configuration not involving the $(2\subset 5)$-brane
that can be obtained by adding a KK monopole to the $2\perp 2\perp
2$ brane. It provides a new embedding of the extreme four dimensional
dyonic black hole with finite horizon area. This solution is described
by
\eqn
\arr{l}
\arr{ll}
ds^{2}= & \left(H_{1}H_{2}H_{3}\right)^{\frac{1}{3}}
\left[ -\left(H_{1}H_{2}H_{3}\right)^{-1}dt^{2}+
H_{1}^{-1}\left( dy_{1}^{2}+dy_{2}^{2}\right) \right.
\\\\ & +H_{2}^{-1}\left( dy_{3}^{2}+dy_{4}^{2}\right)
+H_{3}^{-1}\left( dy_{5}^{2}+dy_{6}^{2}\right) 
\\\\ & \left. +H_{4}^{-1}\left(dy_{7}\pm \a_{4}\cos{\theta}
d\phi\right)^{2}+H_{4}\left( dr^{2}+r^{2}d\o_{2}\right) \right] ,
\earr
\\\\  **{\cal F}={\displaystyle \sum_{i=1}^{3}}
Q_{i}\left(\e_{2}\wedge\eta_{i}\right) .
\earr
\eeqn

This background admits the killing spinor field
\eqn
\e =\left( H_{1}H_{2}H_{3}\right)^{-\frac{1}{6}}\e_{0},
\eeqn
where
\eqn
\G_{012}\e_{0}=\mp \e_{0},\ \
\G_{034}\e_{0}=\mp \e_{0},\ \
\G_{056}\e_{0}=\mp \e_{0}.\ \
\eeqn
The supersymmetry breaking condition due to the KK monopole is
$\G_{78910}\e_{0}=\pm \e_{0}$. Consider first the upper sign choice,
i.e. positive KK charge, this condition can be obtained from (3.15) if
there are three positively or one positively and two negatively
charged 2-branes. If this is the case the solution preserves
$\frac{1}{8}$ of the maximal supersymmetry. Otherwise the KK
monopole supersymmetry breaking condition is incompactible with (3.15)
and the solution breaks all supersymmetries. Similar comments apply for
negative KK charge. An analogous situation occurs for the other
embeddings of the extreme four dimensional dyonic black hole with
finite horizon area. 

Compactifying the solution (3.13) along the
$y_{7}$ direction we obtain the $(2\perp 2\perp 2)\Vert 6$ brane of
type IIA superstring theory 
\eqn
\arr{l}
\arr{ll}
ds_{10}^{2}= & \left(H_{1}H_{2}H_{3}H_{4}\right)^{\frac{1}{2}} \left[
-\left(H_{1}H_{2}H_{3}H_{4}\right)^{-1}dt^{2}+
\left(H_{1}H_{4}\right)^{-1}\left( dy_{1}^{2}+dy_{2}^{2}\right) \right.
\\\\ &
\left. +\left(H_{2}H_{4}\right)^{-1}\left( dy_{3}^{2}+dy_{4}^{2}\right)
  +\left(H_{3}H_{4}\right)^{-1}\left( dy_{5}^{2}+dy_{6}^{2}\right)
  +dx^{j}dx_{j} \right] ,
\earr
\\\\ e^{2\phi}=\left(H_{1}H_{2}H_{3}\right)^{\frac{1}{2}}
H_{4}^{-\frac{3}{2}},
\ \  *{\cal F}_{4}={\displaystyle \sum_{i=1}^{3}}
Q_{i}\left(\e_{2}\wedge\eta _{i}\right) ,
\ \ {\cal F}_{2}=P_{4}\e _{2}.
\earr
\eeqn
Applying T-duality along one direction of each 2-brane we get the 
$3\perp 3\perp 3\perp 3$ brane of type IIB superstring
theory \cite{kt}.

\subsubsection{$2\subset 5 + boost + KK\ monopole$ brane}\

It is straightforward by applying the composite M-brane rule (iv) to
obtain the brane solution $2\subset 5 + boost + KK\ monopole$. It is
given by
\eqn
\arr{l}
\arr{ll}
ds^{2}= & \left(H_{1}\tilde{H}_{1}\right)^{\frac{1}{3}}
\left[ H_{1}^{-1}\left( -dt^{2}+dy_{1}^{2}+\frac{\a_{2}}{r}(dy_{1}\mp
  dt)^{2}+dy_{2}^{2}\right)\right.
\\\\ & +\tilde{H}_{1}^{-1}\left( dy_{3}^{2}+dy_{4}^{2}+
dy_{5}^{2}\right)+dy_{6}^{2}
\\\\ & \left. +H_{3}^{-1}\left(dy_{7}\pm \a_{3}\cos{\theta}
d\phi\right)^{2}+H_{3}\left( dr^{2}+r^{2}d\o_{2}\right) \right] ,
\earr
\\\\ {\cal F}={\cal F}_{(2,1)}+{\cal F}_{(5,1)}
-\frac{\a_{1}\sin{2\zeta_{1}}}{2}
\frac{\tilde{H}_{1}^{-2}}{r^{2}}(dr\wedge \xi_{1}),
\earr
\eeqn
where $*{\cal F}_{(2,1)}=Q_{1}(\e_{2}\wedge\eta_{1})$ and
${\cal F}_{(5,1)}=P_{1}(\mu_{1}\wedge\e_{2})$.

This background preserves $\frac{1}{8}$ of the maximal
supersymmetry. The corresponding Killing spinor field is
\eqn
\e =H_{2}^{-\frac{1}{4}}\left( H_{1}\tilde{H}_{1}\right)
^{-\frac{1}{6}}\left[ \left(\tilde{H}_{1}^{\frac{1}{2}}\pm H_{1}
^{\frac{1}{2}}\cos{\zeta_{1}}\right)^{\frac{1}{2}}+\left(\tilde{H}_{1}
^{\frac{1}{2}}\mp H_{1}^{\frac{1}{2}}\cos{\zeta_{1}}\right)
^{\frac{1}{2}}\gamma \right] \e_{0},
\eeqn
where
\eqn
\G_{012} \gamma \e_{0}=\mp \e_{0},\ \ 
\G_{01}\e_{0}=\mp \e_{0},\ \ 
\G_{78910}\e_{0}=\pm \e_{0},
\eeqn
with $\gamma=\G_{012}\G_{678910}$.

Dimensional reduction along the monopole direction $y_{7}$ yields the
$((2\subset 5)+boost)\Vert 6$ brane solution of type IIA superstring
theory. This solution is described by
\eqn
\arr{l}
\arr{ll}
ds_{10}^{2}= & \left( H_{1}\tilde{H}_{1}{H}_{3}\right)^{\frac{1}{2}}
\left[ \left( H_{1}H_{3}\right)^{-1}\left(
  -dt^{2}+dy_{1}^{2}+\frac{\a_{2}}{r}(dt\mp dy_{1})^{2}+dy_{2}^{2}
\right) \right.
\\\\ & \left. +\left(\tilde{H}_{1}H_{3}\right)^{-1}\left(
dy_{3}^{2}+dy_{4}^{2}+dy_{5}^{2}\right) +H_{3}^{-1}dy_{6}^{2}
+dx^{j}dx_{j}\right],
\earr
\\\\
e^{2\phi}=\left( H_{1}\tilde{H}_{1}\right)^{\frac{1}{2}}
H_{3}^{-\frac{3}{2}},\ \ 
{\cal H}=P_{1}(\mu_{1}\wedge \e_{2})\ ,\ \ 
{\cal F}_{2}=P_{3}\e_{2}\ ,\ \ 
\\\\
{\cal F}_{4}={\cal F}_{(2,1)}
-\frac{\a_{1}\sin{2\zeta_{1}}}{2}
\frac{\tilde{H}_{1}^{-2}}{r^{2}}(dr\wedge \xi_{1}).
\earr
\eeqn
A further reduction yields a $((1\subset 4)\perp 0)\Vert 5$ brane
solution of nine dimensional type IIA superstring theory. We could
have reduced the solution (3.17) along the boost direction to obtain
the $(1\subset 4)\perp 0+KK\ monopole$ brane of type IIA superstring
theory. A further reduction along the monopole direction yields a
$((1\subset 4)\perp 0)\Vert 5$ brane solution of the corresponding
nine dimensional theory. Note that this brane solution is not the same
as the previous one as the basic branes charges arise from different
fields strength.

\section{Conclusion}\

\begin{table}
\caption{Composite M-brane scan}
\bigskip
\begin{tabular}{|c|c|c|c|c|} \hline
 & Composite M-brane & SUSY & $p$ & $D=11-p$  \\ \hline \hline
$N=4$        & $2\perp 2\perp 5\perp 5$                          
& $1/8,0$ & $7$ & $\ \ 4$ $\dagger$ \\ \hline
$N=3$        & $(2\subset 5)\perp (2\subset 5)\perp (2\subset 5)$ 
& $1/8$ & $7$ & $4$   \\ 
             & $2\perp 2\perp 2$                                  
& $1/8$ & $6$ &  $\ \ 5$ $\ddagger$ \\ \hline
$N=3$       & $5\perp 5\perp 5 + boost$                          
& $1/8,0$ & $7$ & $\ \ 4$ $\dagger$ \\
+KK charges   & $2\perp 2\perp 2 + KK \ monopole$                  
& $1/8,0$ & $7$ & $\ \ 4$ $\dagger$ \\ \hline
$N=2$        & $(2\subset 5)\perp (2\subset 5)$                   
& $1/4$ & $7$ & $4$   \\ 
             & $(2\subset 5)\perp 2$                              
& $1/4$ & $6$ & $5$   \\
             & $2\perp 2$                                         
& $1/4$ & $4$ & $7$   \\ \hline
$N=2$       & $(2\subset 5)\perp 5 + boost$                      
& $1/8$ & $7$ & $4$   \\
+KK charges   & $(2\subset 5)\perp 2 + KK \ monopole$              
& $1/8$ & $7$ & $4$   \\
             & $2\perp 5 + boost + KK \ monopole$                 
& $1/8,0$ & $7$ & $\ \ 4$ $\dagger$ \\
             & $2\perp 5 + boost$                                 
& $1/8$ & $6$ & $\ \ 5$ $\ddagger$ \\ \hline
$N=1$        & $2\subset 5$                                       
& $1/2$ & $5$ & $6$   \\ 
             & $2$                                                
& $1/2$ & $2$ & $9$   \\ \hline 
$N=1$       & $(2\subset 5)+boost + KK \ monopole$               
& $1/8$ & $7$ & $4$   \\       
+KK charges   & $(2\subset 5) + boost$                             
& $1/4$ & $5$ & $6$   \\
             & $2+boost$                                          
& $1/4$ & $2$ & $9$   \\ \hline
KK charges   & $boost+KK \ monopole $                                   
& $1/4$ & -   & $4$   \\
             & $boost$
& $1/2$ & -   & $10$  \\ \hline 
\end{tabular}
\end{table}

We have found new intersecting M-brane configurations that follow from
the composite M-brane rules stated in section 2. These new
supersymmetric solutions are built by intersecting the 2, 5 and
$2\subset 5 $ branes and, in some cases, by adding KK charges. 
In table 1 we present the composite M-brane scan resulting
from intersecting $N$ of the 2, 5 and $2\subset 5 $ branes. These
solutions may be seen as anisotropic $p$-branes. The
corresponding amount of preserved supersymmetry is also shown. There
are configurations that are not presented but can be obtained from
other configurations in the table, e.g. the basic 5-brane can be
obtain from the $(2\subset 5)$-brane. By allowing the harmonic
functions to depend on just some coordinates of the overall transverse
space, all configurations with
$p<7$ can be made a r-brane with $p<r\le 7$. Dimensional reduction
along the branes spatial directions yields black hole solutions in
$D=11-p$ dimensions (for $p<7$ we can have $4\le D\le 11-p$). The
solutions marked with $\dagger$
and $\ddagger$ are respectively, embeddings of the four $[23-25]$ and
five \cite{ts2} dimensional dyonic black holes with finite horizon
area. Note that the latter always yields the former by superposing a KK
monopole.

It seems interesting to study the supersymmetric brane solutions of
string theories that arrive by dimensional reduction of these M-branes
and further duality transformations $[11,13,17-19]$. This
could provide a set of composite rules derived from the rather simple
rules in M-theory. 

All solutions that have been considered can be generalised to the 
corresponding black configurations, providing an opportunity of doing 
thermodynamics by using the corresponding M-theory rules
$[34-36]$.
\

\section*{Acknowledgements}\

The author is grateful to M.J.Perry for introducing him into the
M-theory subject and acknowledges the financial support of JNICT
(Portugal) under program PRAXIS XXI.

\end{document}